# Hard Choices and Hard Limits for Artificial Intelligence

Bryce Goodman
Department of Philosophy
University of Oxford
United Kingdom
bwgoodman@gmail.com

## ABSTRACT

Artificial intelligence (AI) is supposed to help us make better choices [1]. Some of these choices are small, like what route to take to work [27], or what music to listen to [64]. Others are big, like what treatment to administer for a disease [74] or how long to sentence someone for a crime [2]. If AI can assist with these big decisions, we might think it can also help with hard choices, cases where alternatives are neither better, worse nor equal but on a par [15]. The aim of this paper, however, is to show that this view is mistaken: the fact of parity shows that there are hard limits on AI in decision making and choices that AI cannot, and should not, resolve.

## KEYWORDS

Value alignment; fairness; hard choices; AI ethics

## 1 Introduction

Artificial intelligence (AI) is supposed to help us make better choices [1]. Some of these choices are small, like what route to take to work [27], or what music to listen to [64]. Others are big, like what treatment to administer for a disease [74] or how long to sentence someone for a crime [2]. If AI can assist with these big decisions, we might think it can also help with hard choices. The aim of this paper, however, is to show that this view is mistaken.

The stakes are as follows. We live in a world where AI is increasingly used to make decisions with significant ethical consequences. In many cases, AI predictions already outperform human judgement, and this trend is poised to continue. In the coming years, many of the decisions currently made by people will – and *should be* – delegated to AI. The question is whether there are theoretical limits on AI in decision making and choices that AI cannot, and should not, resolve.

This paper is structured into the following sections:

1. The first seeks to identify what it is that makes a choice hard. In this section, I reconstruct and defend the philosopher Ruth Chang's position that hard choices occur because alternatives are "on a par": one is neither better, worse nor equal to the other and yet the alternatives are comparable.
2. The second situates Chang's conclusions in the context of the utility model of rational choice. This model says that rational agents' choices are governed by a utility *function*[1], and denies the possibility of parity. If we accept Chang's argument, the utility model fails in the case of hard choices.
3. The third discusses the implications for AI writ large. Every application of machine learning requires an objective function, which attempts to compress a complex problem into a singular scalar value. Thinkers like Stuart Russell believe that we can use AI to learn an agent's reward function, and therefore guarantee alignment between AI goals and human values. But the fact of hard choices proves this approach is wrong: human decision making cannot always be represented by an objective function because human values are not always amenable to quantitative representation.
4. The last provides evidence of the limits of AI decision making through a discussion of algorithmic fairness. Fairness is an essentially contested subject, there is no obvious or objectively correct definition: certain conceptions of fairness are not better, worse or equal to one another – they are on a par. The result is that only people, as moral agents, are capable of committing to a particular conception of fairness, which can then be instantiated in an AI system. There are indeed theoretical limits on AI in decision making.

Though my assessment of AI and hard choices is critical, its conclusion is positive: hard choices don't show that AI is unimportant or uninteresting, but that human agency is *more* important, and more complex, than we might have thought.[2]

## 2 What makes a choice hard?

The following section provides an overview and defense of Ruth Chang's theory of hard choices. Readers already familiar with Chang's work should feel free to skim, as should those who are less philosophically inclined and/or willing to accept Chang's conclusions without argument.

---

[1] Utility function, reward function and objective function are each used in different disciplines (e.g. economics, machine learning, reinforcement learning) and, for our intents and purposes, interchangeable. What matters is that they are all *functions*, "binary relation[s] between two sets that associate to each element of the first set exactly one element of the second set" [81].

[2] For a different, though obviously related, attempt to cash out the implications of Chang's argument for AI, see [23].

Life is filled with choices. Some are hard. Should I pursue a career that brings personal fulfillment, or financial security? Should I live close to my parents, or set off on my own? Do I favor the Beatles, or the Rolling Stones?

But a decision being *big* is not the same as a decision being *hard*. Big choices are big because of their consequences: deciding where we live, who we marry, or whether we undergo a medical procedure are all big choices because the outcome of the choice will have a big impact on our lives. But big choices are not always hard and hard choices are not always big [15:27]. It is a big choice to marry the person we love, but not necessarily a hard choice. On the other hand, deciding whether to spend your holiday in Italy or Ethiopia is a choice that may be small, in terms of overall life consequences, yet hard, because it lacks a clearly correct answer.

It's also worth distinguishing between epistemically and metaphysically hard choices. Some choices clearly have a right answer, even if it is unclear what the right answer is, e.g. will a particular medical treatment improve a patient's condition? In other cases, it is not clear that there is a right answer, but one must choose anyway, e.g. what is the best way to allocate limited medical resources during a pandemic? The latter are core cases of hard choices, not the former.

Chang [15:1] proposes the following gloss of situations that involve hard choices:
1. One alternative is better in some relevant respects
2. The other alternative is better in other relevant respects, and yet
3. Neither seems to be at least as good as the other overall, that is, in all relevant respects

If we think of rational agency as consisting in the ability to recognize and respond to reasons[3], a hard choice occurs when "the agent's reasons for going one way as opposed to the other have, in some sense, 'run out'"[15:1]. Chang's claim is that reasons run out because of a structural feature of hard choices, namely that the alternatives under consideration are neither better, worse nor equal – they are on a par.

Chang defends her position against three alternatives. The first is an argument for ignorance or uncertainty: "what makes a choice hard is that we are ignorant or uncertain of the normative or nonnormative factors that are relevant to making the choice" [15:3]. An agent simply does not know enough about each option to decide: she may be unsure of how a choice will play out, or ignorant about what she actually values.

Deciding how to invest one's money is difficult. But it is not difficult because we don't know what we want. It is difficult because we don't know what choice is most likely to result in what we want. If we knew that investing in stock A would bring us twice the return of stock B, we would buy it. The choice is difficult just because we don't know how things are going to turn out – we are ignorant of nonnormative factors relevant to making the choice.

On the other hand, deciding whether to donate to Save the Children or Greenpeace may be a difficult choice because we are not really sure what cause we should care most about: we are genuinely unsure of what matters more, saving a child's life or protecting the world's oceans. This is also a case of ignorance, but this time we are uncertain about the normative factors relevant to making the choice.

To determine whether a choice's difficulty arises in virtue of ignorance or uncertainty, we can pose the counterfactual: "Would it still be a difficult choice if we had perfect knowledge of all the normative and nonnormative factors relevant to making the choice?" If the argument for ignorance is correct, then in the case of hard choices the answer must always be no.[4]

Chang argues that ignorance cannot be *the* cause hard choices because there are cases where the chooser has first person authority over the properties relevant to the choice, but the choice is hard nevertheless.[5] She offers us the example of choosing between lemon sorbet and apple pie where the only relevant aspect is tastiness to the chooser.[6] Let us suppose we are able to sample each so we know precisely how each dessert tastes. It is plausible that neither tastes better overall: both are incredibly delicious in their own right. And, Chang argues, it is also possible that they do not taste equally good. If apple pie and lemon sorbet *did* taste equally good, adding a small improvement to one should make the choice obvious. But, she argues, this is not so. We may find that apple pie tastes better with whipped cream than without. If we are stuck choosing between apple pie and lemon sorbet, it is not clear that adding whipped cream to the pie will resolve the issue. However, if apple pie and lemon sorbet *were* equal, a small improvement to one should lead us to rationally prefer that option. Chang [15:5–6] calls this the "small improvement argument":
1. A is neither better nor worse than B with respect to V (for example, Apple pie is neither better nor worse than lemon sorbet with respect to tastiness).
2. A+ is better than A with respect to V: (Apple pie with whipped cream is better than apple pie without whipped cream with respect to tastiness).
3. A+ is not better than B with respect to V: (Apple pie with whipped cream is not better than lemon sorbet with respect to tastiness).

---

[3] Chang [16] refers to this as the "passivist view" of rational agency (in contrast to her own "activist view", discussed later in our paper) and finds its roots in the works of Susan Wolf [78], Joseph Raz [55], Thomas Scanlon [63], Derek Parfit [54], Jonathan Dancy [22], Jonathan Skorupski [68] and, more distantly, Hobbes [38:45], who wrote that "when a man *reasons*, he does nothing else but conceive a sum total from addition of parcels for REASON…is nothing but reckoning".

[4] To be clear, a choice can be intuitively "hard" even if we know what choice we need to make – you may know that the right thing is to end a relationship, but nevertheless struggle with the decision. This sort of difficulty, however, does not pertain to knowing what choice to make, but rather the significance and/or consequences of making that choice.

[5] In other words, the chooser has perfect knowledge of the relevant normative and nonnormative factors, but the choice is still hard.

[6] If this example seems overly trivial, we might suppose that this will be the chooser's last meal, or swap in more somber stakes, e.g. Sophie's choice of which child to save.

4. A and B are not equally good with respect to V: (If they were, apple pie with whipped cream would taste better than lemon sorbet but in premise 3 we have assumed it does not).
5. Therefore, A is not better than, worse than, or equally as good as B with respect to V

The dessert example illustrates the possibility of cases where "none of the three usual comparative relations, 'better than', 'worse than', and 'equally good'" holds even though the chooser has perfect knowledge of the relevant aspects of the choice [15:4]. Consequently, ignorance alone cannot explain all cases of hard choices that fit our gloss.

The second objection Chang rejects holds that hard choices occur when the alternatives are incommensurable. To say that alternatives are incommensurable is to say that they lack a common measure [56:1,15:6]. Weight and height are, in this sense, incommensurable properties: it is incoherent to compare an object's height with another object's weight. On this view, apple pie and lemon sorbet are incommensurable because tastiness is not the sort of property that can be represented by units: we cannot say that apple pie is 2.4 times as tasty as lemon sorbet. So, one might argue, hard choices arise because we lack the ability to quantitively compare alternatives.

This argument, however, conflates incommensurability with incomparability. We may not be able to quantify the difference in taste between apple pie and lemon sorbet, but we can nevertheless compare the two. Alternatives that cannot be related cardinally may still be related ordinally: we regularly engage in rational comparisons of incommensurable alternatives by ranking one above the other.

Furthermore, incommensurability between alternatives does not necessarily mean that a choice is hard. There may be no unit for measuring how much enjoyment we get from watching television or falling in love, but this does not mean we would struggle to choose one experience over the other. As Chang [15:7] writes, "since it is compatible with alternatives being incommensurable that one of them is better than the other with respect to [the relevant property], the choice between incommensurables could be easy: choose the better alternative. Thus, incommensurability is not what makes a choice hard".

A third position considered by Chang [15:10] is that a choice is hard when the options "cannot be compared with respect to what matters in the choice". On this view, whenever there is no basis for comparing two options, there is no basis for a rational choice [56]. If incomparability is what makes choices hard than choosing one alternative over another can never be rational. Chang's three arguments against this position essentially boil down to the observation that we can, in fact, make rational decisions even in the case of hard choices: "incomparabilists cannot make sense of the relatively common fact that sometimes in a hard choice the agent rationally concludes that she has most reason to choose one rather than the other of the alternatives" [15:9]. Intuitively, if we think lemon sorbet and apple pie *can* be compared in terms of their tastiness even if their tastiness is incommensurable, we must reject the view that hard choices arise only when and because alternatives are incomparable.[7]

Our discussion thus far has concerned Chang's argument that neither ignorance, incommensurability nor incomparability are the root cause of hard choices. Before proceeding to evaluate Chang's positive claims, it is worth situating her view relative to the classical assumptions of rational choice and expected utility theory, which provide the basis for what I will call the "utility model" of rational choice [51].[8]

## 3 Hard Choices and the Utility Model

According to the utility model of rational choice, "any rational agent can be described as having a utility function that assigns a real number representing the desirability of being in any particular world state" [59:7–8]. In particular, expected utility theory says that a rational decision maker chooses between alternatives by comparing their weighted expected utility values, which is a function of utility value and probability [50]. When the choice takes place under certain conditions, as in the dessert example, the utility value alone is what determines the choice.[9]

The utility model, and its assumptions about how people make choices, plays a prominent role in economics [30], public policy [35,39], ethics [47], psychology [25,42] and artificial intelligence [33,31,60]. As a descriptive theory, it offers a tool for understanding and predicting why people choose what they do: "it looks at people's choices (e.g. how much money they've saved, what car they bought), and tries to 'rationalize' those choices, that is, figure out whether the choices are compatible with optimization and, if so, what the choices imply about the agent's preferences" [45:7]. When we rationalize an agent's choices we are trying to create an abstract representation of her preferences that will make sense of the decisions she made and, insofar as she acts rationally, predict what decisions she is likely to make.[10]

Under the utility model, an agent chooses between alternatives in accordance with a set of preferences and her choice is rational insofar as those preferences are governed by a set of logical relations. In particular, it is assumed that an agent's preferences are both *transitive* and *complete* [73:66–67]. Transitivity means that if a rational agent prefers A to B and B to C, she will also prefer A to C. Completeness means that all of an agent's options can be related in one of three ways:
1. She prefers A to B
2. She prefers B to A

---

[7] This is a very short and incomplete version of an argument Chang [14] offers in favor of *comparativism*, which says (among things) that we can rationally choose between incomparable alternatives. This claim is not central to our argument and so is only cited for completeness.
[8] There are, of course, lively debates within the field of decision theory that I am passing over, such as whether the expected utility model is a normative or positive theory.
[9] Inverse reinforcement learning, which we will discuss later, assumes what is sometimes called "Boltzmann (ir)rationality", according to which humans are "noisily rational" and will choose an option with a probability proportional to *e* to the power of its utility [79,66].
[10] As a normative theory, the standard model would also tell us what decisions she *should* make.

3. She is indifferent between A and B

This trichotomy should be familiar – it exhaustively describes the way in which any two quantities can be related. For example, an object may weigh more, less, or the same as another. There is no fourth option.

If we accept this model, we can make a large number of inferences from a small number of choices. For example, if Alice chooses apple pie over lemon sorbet and we assume that tastiness is the only factor relevant to her choice, the utility model says we can infer that she thinks apple pie is either better than or equal to lemon sorbet with respect to tastiness. Further, if we learn that she prefers apple pie with whipped cream to apple pie, we can reliably predict that, insofar as she is rational, she will *always* tend towards choosing apple pie with whipped cream over lemon sorbet.

If Chang's theory of hard choices is true, however, it poses a particular challenge for the utility model of rational choice. Chang rejects the trichotomy thesis and introduces a fourth option: A and B may be *on a par* [12]. When alternatives are on a par they are comparable yet incommensurable: we can rationally choose one over the other *even though* we lack a common metric for comparison. And, she argues, it is precisely in these cases that we end up facing a hard choice.

If Chang's account is accurate, the trichotomy assumed by the utility model provides an incomplete vocabulary to describe rational choice. We must introduce a fourth relationship in which A is neither better, worse nor equal to B. Further, parity is not transitive: if A is on a par with B and B is on a par with C it does not follow that A must be on a par with C. Alice may think apple pie with whipped cream is on a par with lemon sorbet and that lemon sorbet is on a par with apple pie but still prefer apple pie with whipped cream over plain old apple pie [15:21]. The upshot is that, if we accept parity as a relationship between alternatives, we are limited in what we can infer about a rational agent based on her choices alone: if Alice picks apple pie over lemon sorbet, we cannot conclude she thinks that apple pie is better or equally good. She may, instead, think that they are on a par.

The utility model may still be able to accurately describe and predict how agents make decisions in many situations, in particular those situations that involve a choice between alternatives that *can* be quantitively compared (e.g. money). If we could reliably distinguish cases where an agent faces a hard choice we could appropriately circumscribe the model. However, the fact that hard choices are caused by structural, rather than substantive, features of a choice means that there is no obvious way to determine, *a priori*, when hard choices will arise for an agent. Furthermore, because hard choices are structural, they might arise far more often than we might think – as we have already seen, a hard choice does not have to be big.

So, what are we to do when we face a hard choice? How can we make a rational decision when alternatives are on a par? Chang's answer begins by distinguishing between reasons that rest upon facts external to the actor (*given* reasons) and those that exist in virtue of the actor's will (*will-based* reasons). According to Chang [15:23], "when our given reasons are on a par, will-based reasons may step in…by putting your will behind a feature of an option—by standing for it—*you* can be that in virtue of which something is a will-based reason for choosing that option". In hard choices, it is an agent's will, rather than any external condition, that serves as the determinative source of reasons.

Chang's argument is at once intuitive and surprising. It is intuitive that the exercise of our will – the act of deciding to do something – may be a source of reasons. After all, when we commit to a cause or a person, that commitment grounds reasons (e.g. to spend our time with them, to protect them, etc.). But it is surprising that rational agency – the ability to decide based on reasons – may require the ability to not merely recognize but also *generate* reasons [16].

## 4 Hard Choices and the Limits of AI

We now turn our attention to how hard choices should affect how we think about AI. Although many of the conclusions we reach will be applicable to AI more generally, our analysis will focus on machine learning, a specific sub-field of AI concerned with algorithms that "learn" from data. But first, it is worth briefly saying something about what AI is, and what it is not.

The language of AI is math: if a problem cannot be stated mathematically, it is not suitable for AI. More specifically, we cannot use AI unless we can provide a quantitative description of what it is we want to optimize. In machine learning, an algorithm learns to map a set of inputs to outputs. A map learned by a machine learning algorithm is called a model.[11] Models are produced through a process of training. In *supervised learning*, each training iteration involves measuring the success of the model at mapping inputs to outputs and then adjusting the model. The difference between a model's output (prediction) and the actual output (reality) is known as error. The goal of training is to reduce model error. The function used to track how well the model is doing is the objective function: this function's input is the model's results and the output is a scalar value. When the function outputs higher values for more accurate results, we call it a reward function. In the opposite case, we call it the cost, loss or error function [9:82]. If the algorithm is intended to optimize an agent's utility (e.g. evaluate and recommend the best alternative), the objective (or reward) function is equivalent to the utility function discussed in the previous section.

A well specified objective function is critical to ensure alignment between what an AI system does and what we want it to do. This is because the objective function "reduces all the various good and bad aspects of a possibly complex system down to a single number, a scalar value, which allows candidate solutions to be ranked and compared" [57:155]. As a practical matter, our ability to build an AI system that meets our goals will only ever be as good as our ability to translate those goals into a single metric.

---

[11] Deep learning is a subfield of machine learning that builds models using neural networks [9].

The same is true for another popular field of machine learning: *reinforcement learning*. In the context of AI[12], reinforcement learning is concerned with artificial agents or learners "learning what to do—how to map situations to actions—so as to maximize a numerical reward signal. The learner is not told which actions to take, but instead must discover which actions yield the most reward by trying them" [71:2]. The "numerical reward signal" is specified by a reward function, which typically assigns different values to states brought about by an agent's actions [59:8].[13] More generally, reinforcement learning adopts what Rich Sutton calls the *reward hypothesis*: "all of what we mean by goals and purposes can be well thought of as maximization of the expected value of the cumulative sum of a received scalar signal (reward)" [70].

The upshot is that both supervised and reinforcement learning – the two dominate modes underpinning AI today – depend upon the ability to represent decisions and their results in quantitative, scalar terms.

If Chang's theory is correct, it places clear limits on the reward hypothesis. More specifically, it says that we should reject the reward hypothesis whenever we encounter a hard choice, since in this scenario goals and purposes *cannot* be represented in terms of a scalar signal. Furthermore, it gives us criteria for deducing what scenarios or sorts of problems will elude AI solutions on philosophical, and not merely technical, grounds.

What does this mean for AI? It means that certain claims about the future of AI capabilities are *a priori* false. Demis Hassabis, co-founder and CEO of DeepMind – arguably the most advanced AI company in the world – has stated that DeepMind's goal is "to solve intelligence" and then use that intelligence "to solve everything else" [80]. But if Chang's thesis is true, solving intelligence won't solve everything else. In particular, it won't solve problems that involve hard choices.

Chang's theory also calls into question the merits of certain solutions to the so-called "alignment problem" in AI: the problem of ensuring agreement between AI objectives and human values [19].[14] One category of solutions is based upon *inverse reinforcement learning*, which attempts to solve the following [52]:

Given
1. Measurements of an agent's behavior over time, in a variety of circumstances;
2. if needed, measurements of the sensory inputs to that agent; and
3. if available, a model of the environment

Determine the reward function being optimized

Put simply, the inverse reinforcement learning problem is the problem of inferring from an agent's behavior and circumstances the reward or utility function that is guiding her choices.[15] Various forms of inverse reinforcement learning have been offered as a solution to the alignment problem [33,32]. For example, Stewart Russell [59:7,60] proposes three principles to guide the design of "provably beneficial" AI:

1. The machine's purpose is to maximize the realization of human values.
2. The machine is initially uncertain about what those human values are.
3. Machines can learn about human values by observing the choices that we humans make.

Russell calls this "cooperative inverse reinforcement learning", and the basic idea is that AI can learn what someone values – her reward function – through repeat interactions and observations.[16]

However, if Chang is correct, humans do not have anything like a reward function. Sometimes, perhaps oftentimes, goals and purposes simply cannot be represented as the maximization of the expected value of a scalar reward [75]. Russell's approach fails, not because AI cannot come to learn what we value per se, but because what we value is partly determined by *internally generated reasons* that are not revealed by externally observable behavior [13].[17]

More generally, if we accept the possibility of parity, we allow for the possibility that a rational agent's preferences may be *underdetermined* up and until she makes a choice.[18] This conclusion is supported not only by Chang's philosophical arguments, but empirical evidence. As the neuropsychologists Benjamin Hayden and Yael Niv [36:7] put it, "value doesn't sit in the brain waiting to be used; rather, preference is a complex and active process that takes place at the time the decision is made". The prevalence of *ad hoc* preference formation is also evidenced by the work of behavioral scientists Sarah Lichtenstein and Paul Slovic [46:1], who write:

---

[12] Outside of AI, reinforcement learning "has furnished the canonical computational framework for understanding value-guided decision-making in humans and other primates, and has been widely used to explain how different brain regions participate in valuation and choice" [41:856].

[13] The issue of how to properly specify reward functions plays a prominent role in discussions about AI safety. If we want to train a robot to sweep our floors, for example, we may design a reward function that assigns positive value to the action "cleaning" or state "clean floors". If the reward function assigns positive value every time a floor goes from dirty to clean, our robot may get stuck in an endless cycle of cleaning, dumping and cleaning again, since this would produce a greater reward than simply cleaning the floors once. This behavior is known as "reward hacking" because the AI agent has "hacked" the reward function [20].

[14] This same concern is also sometimes referred to as the "control problem": "how to ensure that systems with an arbitrarily high degree of intelligence remain strictly under human control" [59:3]. See also [10].

[15] This is, more or less, an operationalization of economist Paul A. Samuelson's [61] "Consumption theory in terms of revealed preference". For two canonical critiques, see [65,76].

[16] In particular, Russel [59:8] states that "to the extent that *objectives* can be *defined* concisely by specifying reward functions, *behavior* can be *explained* concisely by inferring reward functions".

[17] This conclusion is in accord with Stuart Armstrong and Sören Mindermann [5:2], who argue that "although current [inverse reinforcement learning] methods can perform well on many well-specified problems, they are fundamentally and philosophically incapable of establishing a 'reasonable' reward function for the human, no matter how powerful they become." Where we differ, however, is over *why* such methods fail: they argue that "normative assumptions" need to be introduced to constrain the space of potential reward functions assignable to observed behavior, whereas we have argued that an agent's preferences may be underdetermined up to and until a choice is made.

[18] A recent study by Silver et al [67] found "that infants experienced choice-induced preference change similar to adults'…hence, choice shapes preferences—even without extensive experience making decisions and without a well-developed self-concept."

One of the main themes that has emerged from behavioral decision research during the past three decades is the view that people's preferences are often constructed in the process of elicitation. This idea is derived from studies demonstrating that normatively equivalent methods of elicitation (e.g., choice and pricing) give rise to systematically different responses. These preference reversals violate the principle of procedure invariance that is fundamental to all theories of rational choice.

The only way to square this evidence with the idea that humans are rational agents is to accept that a rational agent need not have fully formed preferences prior to the act of choice and that "there is no guarantee of consistency or reliability [of preferences]" [36:7].

The upshot is that, when we 'rationalize' an agent's choices under the utility model, we may be imposing consistency where there is none.[19] Instead, we should recognize that inconsistency is not a sign of irrationality – it is a mark of agency. As Emerson [26:183] reminds us, "a foolish consistency is the hobgoblin of little minds…with consistency a great soul has simply nothing to do. He may as well concern himself with his shadow on the wall".

To recap, all applications of machine learning require an *objective function* that "reduces all the various good and bad aspects of a possibly complex system down to a single number, a scalar value, which allows candidate solutions to be ranked and compared" [57:155]. In particular, reinforcement learning adopts the *reward hypothesis*: "all of what we mean by goals and purposes can be well thought of as maximization of the expected value of the cumulative sum of a received scalar signal (reward)" [70]. However, hard choices are cases where alternatives *cannot be represented* in terms of scalar values: the standard trichotomy– greater than, lesser than and equal to – simply does not apply. Therefore, we should not expect that AI can resolve those choices.

This conclusion has important implications for the relationship between intelligence and agency. Pei Wang [29:31] defines intelligence as "the capacity of a system to adapt to its environment while operating with insufficient knowledge and resources". If we accept this definition, it seems perfectly plausible that machines can satisfy the requirements of being (at least somewhat) intelligent. But agency – the ability to engage in making rational choices – is quite a different matter. Our argument reveals that a rational agent must not only recognize reasons but, when faced with a hard choice, be capable of *generating* reasons.[20]

All of which points to a new, and critical, question: can we rely on AI to generate the sort of reasons required in hard choices? And, perhaps more importantly, should we? We turn to this question in the following section.

## 5 Hard Choices in AI Design: The Case of Fairness

If we can rely upon AI to resolve hard choices, it must follow that we can rely upon AI to resolve hard choices *in its design*. A closer examination reveals that this is not so, nor should we want it to be.

Imagine we are designing an algorithm that will be used to recommend candidates for a job, and we want the algorithm to be fair with regards to a candidate's sex. We immediately face a hard choice: how should we define fairness? For some, fairness depends on everyone being treated the same: we should all be subject to the same rules and processes. For others, what really matters is not the process, but where people end up: if applying the same rules to everyone results in one group having less than another, we shouldn't subject everyone to the same rules. The debate has no *obviously correct*[21] resolution because fairness is what W.B. Gallie [28:168] calls an 'essentially contested concept': "when we examine the different uses of these terms and the characteristic arguments in which they figure we soon see that there is no one clearly definable general use of any of them which can be set up as the correct or standard use".

The mere fact people cannot agree on how a term should be used does not mean it refers to an essentially contested concept: there are lively debates over whether human-induced climate change is real or child vaccinations cause autism, but this is just because certain groups are scientifically misguided. However, in the case of fairness, the fact that we lack a universally accepted conception is not the result of ignorance or obstinance. Rather, insofar as fairness is an essentially contested concept, disputes "are perfectly genuine" and "although not solvable by argument of any kind, are nevertheless sustained by perfectly respectable arguments and evidence" [28:169].

Most of the time, our world is able to accommodate such conceptual debates: art galleries will remain open even if we disagree over the essential nature of art. However, if we are to ensure that an *algorithm's* recommendations are fair, we must be able to first express fairness in mathematical terms. This is where the trouble begins.

Numerous methods have been proposed for making machine learning algorithms 'more fair' [77,40]. Mehrabi et al. [48] provide an overview of more than twenty one different definitions of 'algorithmic fairness' found in machine learning literature, such as:

- Equalized odds – Men and women should have equal rates for true positives and false positives [34,58,49,7].
- Equal opportunity – Men and women should have equal true positive rates [6,34,8].
- Demographic parity – The likelihood of a positive outcome should be the same regardless of whether the person is a man or a woman [21].

---

[19] One is reminded of Nassim Nicholas Taleb's *Bed of Procrustes* [72]: "we humans, facing limits of knowledge, and things we do not observe, the unseen and the unknown, resolve tension by squeezing life and the world into crisp commoditized ideas, reductive categories, specific vocabularies, and prepackaged narratives..."

[20] Sutton [69] notes that John McCarthy, who coined the term "artificial intelligence", believed that "intelligence is the computational part of the ability to achieve goals in the world." This is perfectly commensurate with our argument, which says that rational agency requires an additional capacity: to establish those goals.

[21] The fact there is no obvious correct solution does not mean there are not obviously incorrect solutions (e.g. give all the jobs to men) or that a solution cannot be reached. The point is that fairness, unlike accuracy, is not an objective measure.

- Fairness through awareness – Men and women with similar characteristics should receive similar outcomes [24].
- Fairness through unawareness – The algorithm is fair so long as gender is not explicitly used in the decision-making process [18].

Each conception captures a different intuition about what might constitute fairness. One would hope that all of these intuitions are mutually compatible in practice. But they are not.

Kleinberg et al. [44] show that various conceptions of fairness in machine learning derive from an attempt to satisfy one of three distinct conditions:

- Calibration condition -- That the probability estimates provided by the algorithm should be well calibrated across groups: if the algorithm predicts that x% of women and y% of men are good candidates, then x% of women and y% of men *should be* good candidates.
- Balance for the positive class condition -- That the average score (probability estimate) received by people *recommended* by the algorithm is the same across groups: if women who score x or greater are recommended, then men who score x or greater *should also* be recommended.
- Balance for the negative class condition -- That the average score (probability estimate) received by people *rejected* by the algorithm is the same across groups: if women who score x or lower are rejected, then men who score x or lower *should also* be rejected.

The researchers consider a number of cases and proposed solutions before concluding that "these conditions are in general incompatible with each other…moreover, this incompatibility applies to *approximate* versions of the conditions as well" [44:3].

All of the metrics available for measuring or achieving fairness will, in nearly all cases, result in trade-offs: intuitions about what constitutes a fair decision will be violated by implementing one alternative or another.[22] The upshot is that deciding how to measure fairness is a hard choice: certain alternatives are neither better, worse nor equal to one another with respect to fairness. They are on a par.

So, what are we to do? Chang has already provided an answer: when faced with a hard choice, we have reached the limits of quantitative comparison. Nothing in the world will tell us the correct answer. Instead, we must *commit*. Our commitment can then become a new source of reasons: "when you commit to something…it is your agency, the very activity of committing, that is the source of the reasons you create" [15:24].[23]

Of course, we do not have to commit. We can neglect to "exercise our normative power as rational agents" and drift into one alternative, that is, "intentionally choose it, but in a way that is noncommittal, that does not involve our putting ourselves behind [it]" [15:26]. But there is a cost to drifting. The main character in Camus' [11] *The Stranger* is horrifying precisely because of the way he drifts into becoming a murderer. Hannah Arendt [3:54] observed that "in the general moral denunciation of the Nazi crimes, it is almost always overlooked that the true moral issue did not arise with the behavior of the Nazis but of those who only 'coordinated' themselves and did not act out of conviction". When the ethical stakes are high, drifting is, according to Arendt, not merely lazy, but evil [4].

On the other hand, the ability to exercise our will and form our own reasons *sui generis* is, for thinkers such as Soren Kierkegaard [43], Friedrich Nietzsche [53:219] and Martin Heidegger [37:248], precisely what it means to live authentically. When we decide whether and how we commit, we prove that we are not "machine men with machine minds and machine hearts" [17] and become the authors of our own rationality.

The present example shows us why certain hard choices are *especially* hard, and why they are, and should, be outside the purview of AI. They require that we commit not only to a particular alternative, but also to a particular understanding of what matters *to us*. In such cases, delegating decision making to AI is not only impossible, it is unconscionable.

Some might find this conclusion dispiriting. However, as Chang notes, navigating hard choices "is the point at which we come into our own as self-governing agents" [15:27]. We should not despair that AI cannot resolve hard choices, because hard choices give us the opportunity to decide who we are.

## ACKNOWLEDGMENTS

I would like to thank the Uehiro Centre for Practical Ethics and the Institute for Ethics in AI at the University of Oxford as well as the following for their invaluable feedback: Ruth Chang, Julian Savulescu, Brian Christian, Ritwik Gupta, Larry Sommer McGrath, Thomas Sinclair, Lofred Madzou and Nikita Aggarwal.

---
[22] This result is also known as the "Impossibility Theorem of Machine Fairness" [62].
[23] I suspect that much of the work on algorithmic fairness has had very little impact because the organizations charged with adopting solutions are, in the absence of any explicit legislation, reluctant to commit to a definition of fairness.